\documentclass[12pt]{iopart}
\usepackage{graphicx}
\usepackage{amssymb}
\begin{document}

\title[Exactly solved spin-$1$ Ising--Heisenberg diamond chain]
{Exactly solvable spin-$1$ Ising-Heisenberg diamond chain with the
second-neighbor interaction between nodal spins}

\author{V.V. Hovhannisyan$^{1}$, J. Stre\v{c}ka$^2$, N.S. Ananikian$^{1,3}$}

\address{$^1$ A.I. Alikhanyan National Science Laboratory, 0036 Yerevan, Armenia\\}
\address{$^2$ Institute of Physics, Faculty of Science, P. J. \v{S}af\'{a}rik University, Park Angelinum 9, 040 01, Ko\v{s}ice, Slovak Republic}
\address{$^3$ Applied Mathematics Research Centre, Coventry University, Coventry, CV1 5FB, England, UK\\}

\begin{abstract}

The spin-1 Ising-Heisenberg diamond chain with the second-neighbor
interaction between the nodal spins is rigorously solved using the
transfer-matrix method. Exact results for the ground state,
magnetization process and specific heat are presented and discussed
in particular. It is shown that the further-neighbor interaction
between the nodal spins gives rise to three novel ground states with
a translationally broken symmetry, but at the same time, it does not
increases the total number of intermediate plateaus in a
zero-temperature magnetization curve compared with the simplified
model without this interaction term. The zero-field specific heat
displays interesting thermal dependencies with a single- or
double-peak structure.
\end{abstract}


\section{Introduction}

The spin-1/2 Ising-Heisenberg diamond chain has received
considerable research interest since \v{C}anov\'a \textit{et al}.
\cite{can04,can06} reported on first exact results for this
interesting quantum spin chain. Early exact results for the spin-1/2
Ising-Heisenberg diamond chain have predicted a lot of intriguing
magnetic features such as an intermediate one-third magnetization
plateau or double-peak temperature dependencies of specific heat and
susceptibility \cite{can04,can06}. In spite of a certain
over-simplification, the generalized version of the spin-1/2
Ising-Heisenberg diamond chain  qualitatively reproduces
magnetization, specific heat and susceptibility data reported on the
azurite Cu$_3$(CO$_3$)$_2$(OH)$_2$, which represents the most
prominent experimental realization of the spin-1/2 diamond chain
\cite{kik03,kik04,kik05a,kik05b}. A lot of attention has been
therefore paid to a comprehensive analysis of quantum and thermal
entanglement \cite{ana12,roj12,tor14,fai14,fai15}, correlation
functions \cite{bel13}, Lyapunov exponent \cite{ana13}, zeros of
partition function \cite{ana14}, magnetocaloric effect \cite{dua14},
the influence of asymmetric \cite{val08,lis11}, further-neighbor
\cite{lis14} and four-spin interactions \cite{gal13,gal14}.

Recently, it has been verified that the spin-1 Ising-Heisenberg
diamond chain may display more diverse ground states and
magnetization curves than its spin-1/2 counterpart
\cite{ans14,abg14}. It actually turns out that the magnetization
curve of the spin-1 Ising-Heisenberg diamond chain involves
intermediate plateaus at zero, one-third and two-thirds of the
saturation magnetization even if the relevant ground states do not
have translationally broken symmetry \cite{ans14,abg14}, while the
spin-1/2 Ising-Heisenberg diamond chain may involve those
intermediate plateaus just if the asymmetric,  further-neighbor
\cite{lis14} and/or four-spin interactions \cite{gal13,gal14} break
a translational symmetry of the relevant ground states. In the
present work, we aim to examine the role of the second-neighbor
interaction between the nodal spins on the ground state and
magnetization process of the spin-1 Ising-Heisenberg diamond chain.

The paper is organized as follows. In Sec. \ref{model} we will
introduce the investigated spin-chain model and briefly describe
basic steps of our rigorous calculation. The most interesting
results for the ground state, magnetization process and specific
heat are discussed in detail in Sec. \ref{result}. Finally, some
conclusions and future outlooks are briefly mentioned in Sec.
\ref{conclusion}.

\section{The model and its exact solution}
\label{model}

We consider the spin-$1$ Ising-Heisenberg model on a diamond chain
in a presence of the external magnetic field. The primitive unit
cell of a diamond chain consists of two Heisenberg spins $S_{a,i}$
and $S_{b,i}$, which interact symmetrically via Ising-type interaction
with two nearest-neighbor Ising spins $\mu_{i}$ and $\mu_{i+1}$
(see Fig.~1). The total Hamiltonian of the model under investigation
may be represented as a sum over block Hamiltonians
$\mathcal{H}=\sum_{i=1}^{N}\mathcal{H}_{i}$, where
\begin{eqnarray}
\mathcal{H}_{i} &=&
J[\Delta(S_{a,i}^{x}S_{b,i}^{x}+S_{a,i}^{y}S_{b,i}^{y})+S_{a,i}^{z}S_{b,i}^{z}] +J_{1}\left(S_{a,i}^{z}+S_{b,i}^{z}\right)\left(\mu_{i}+\mu_{i+1}\right) \nonumber \\
&&+J_2
\mu_{i}\mu_{i+1}-H_{\mathrm{H}}\left(S_{a,i}^{z}+S_{b,i}^{z}\right)
-H_{\mathrm{I}}\frac{\mu_{i}+\mu_{i+1}}{2}. \label{1.1}
\end{eqnarray}
In above, $S_{a,i}^{\alpha}$ and $S_{b,i}^{\alpha}$ ($\alpha=x,y,z$)
denote spatial components of the spin-$1$ operators, $\mu_{i} =
\pm1,0$ stands for the Ising spin, $J$ labels the XXZ interaction
between the nearest-neighbor Heisenberg spins, $\Delta$ is a
spatial anisotropy in this interaction, $J_1$ is the Ising
interaction between the nearest-neighbor Ising and Heisenberg spins
and finally, the last two terms $H_{\mathrm{H}}$ and $H_{\mathrm{I}}$ determine Zeeman's energy of the
Heisenberg and Ising spins in a longitudinal magnetic field.

\begin{figure}[!h]
\includegraphics[width=120mm]{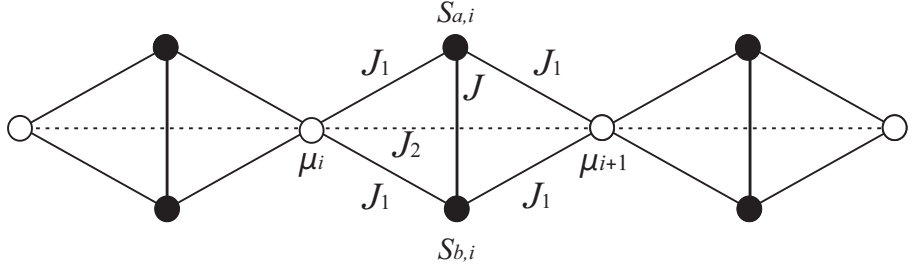}
\caption{The $i$-th block of the spin-1 diamond chain with the Ising
($\mu_{i}$, $\mu_{i+1}$) and Heisenberg ($S_{a,i}$, $S_{b,i}$)
spins.} \label{fig:diamond}
\end{figure}

The important part of our further calculations is based on the
commutation relation between different block Hamiltonians
$[\mathcal{H}_i,\mathcal{H}_j]=0$, which will allow us to partially
factorize the partition function of the model and represent it as a
product over block partition functions
\begin{eqnarray}
Z_N=\sum_{\{\mu_{i}\}}\prod^{N}_{i=1}\mbox{Tr}_{i}\rm e^{-\beta
\mathcal{H}_{i}}, \label{1.2}
\end{eqnarray}
where $\beta=(k_{B}T)^{-1}$, $k_{B}$ is Boltzmann's constant, $T$ is
the absolute temperature, $\sum_{\{\mu_{i}\}}$ marks a summation over
spin states of all Ising spins and $\mbox{Tr}_{i}$ means a trace
over the spin degrees of freedom of two Heisenberg spins from the
$i$-th block. After a straightforward diagonalization of the cell
Hamiltonian (\ref{1.1}), which corresponds to the spin-1 quantum Heisenberg dimer in a magnetic field,
one obtains the following expressions for the respective eigenvalues
\begin{eqnarray}
\mathcal{E}_{1}(\mu_{i},\mu_{i+1})&=& -J+J_2 \mu_i \mu_{i+1}-\frac{H_{\rm{I}}}{2}\mu,\nonumber \\
\mathcal{E}_{2,3}(\mu_{i},\mu_{i+1})&=& \pm J\Delta +J_2 \mu_i \mu_{i+1} - (J_1+\frac{H_{\rm{I}}}{2})\mu +H_{\rm{H}},\nonumber \\
\mathcal{E}_{4,5}(\mu_{i},\mu_{i+1})&=& \pm J\Delta +J_2 \mu_i \mu_{i+1} + (J_1-\frac{H_{\rm{I}}}{2})\mu -H_{\rm{H}},\nonumber \\
\mathcal{E}_{6,7}(\mu_{i},\mu_{i+1})&=&   J + J_2 \mu_i \mu_{i+1} + (\pm2J_1-\frac{H_{\rm{I}}}{2})\mu \mp 2H_{\rm{H}},\nonumber \\
\mathcal{E}_{8,9}(\mu_{i},\mu_{i+1})&=& -\frac{J}{2}(1 \pm \delta)
+J_2 \mu_i \mu_{i+1} - \frac{H_{\rm{I}}}{2}\mu, \label{1.4}
\end{eqnarray}
where $\mu \equiv  \mu_i+\mu_{i+1}$ and $\delta=\sqrt{1+8\Delta^2}$.
Now, one may simply perform a trace over the spin degrees of freedom
of the spin-1 Heisenberg dimers on the right-hand-side of
Eq.~(\ref{1.2}) and the partition function can be consequently
rewritten into the following form
\begin{eqnarray}
Z_N=\sum_{\{\mu_{i}\}}\prod^{N}_{i=1}T_{\mu_{i},\mu_{i+1}} = \mbox{Tr}
\,T^{N}, \label{1.5}
\end{eqnarray}
where the expression $T_{\mu_{i},\mu_{i+1}}$ can be viewed the
standard $3 \times 3$ transfer matrix
\begin{eqnarray}
T_{\mu_{i},\mu_{i+1}} =  \left(
\begin{array}{cccc}
T_{1,1}  & T_{1,0} & T_{1,-1}  \\\
T_{0,1} & T_{0,0} & T_{0,-1} \\\
T_{-1,1}  & T_{-1,0} & T_{-1,-1}
\end{array} \right).
\label{1.7}
\end{eqnarray}
Here, the subscripts $\pm1$ and 0 denote three available spin states of the Ising spins
$\mu_i$ and $\mu_{i+1}$ involved in the transfer matrix (\ref{1.7}),
which has precisely the same form as the transfer matrix of the generalized
spin-1 Blume-Emery-Griffiths chain diagonalized in Refs.~\cite{kri74,kri75,lis13}.
Of course, individual elements of
the transfer matrix (\ref{1.7}) are defined through the formula
\begin{eqnarray}
T_{\mu_{i},\mu_{i+1}}=\mbox{Tr}_{i}\rm e^{-\beta
\mathcal{H}_{i}}=\sum_{n=1}^{9}\rm
e^{-\beta\mathcal{E}_{n}(\mu_{i},\mu_{i+1})}, \label{1.6}
\end{eqnarray}
which includes the set of eigenvalues (\ref{1.4}) for the spin-1
quantum Heisenberg dimer and can be rewritten as
\begin{eqnarray}
T_{\mu_{i},\mu_{i+1}} = \mathrm{exp}(-\beta J_2 \mu_i \mu_{i+1}+\beta
H_{\rm{I}} \mu /2)[2 \mathrm{exp}(\beta J/2)\mathrm{cosh}(\beta J
\delta /2) +\mathrm{exp}(\beta J)  \nonumber
\\ +4 \mathrm{cosh}( \beta J
\Delta) \mathrm{cosh}(\beta J_1 \mu -\beta H_{\rm{H}})
+2\mathrm{exp}(-\beta J)\mathrm{cosh}(2\beta J_1 \mu -2\beta
H_{\rm{H}})]. \label{1.7a}
\end{eqnarray}
With regard to Eq.~(\ref{1.5}), the partition function of the spin-1
Ising-Heisenberg diamond chain can be expressed through three
eigenvalues of the transfer matrix \cite{kri74,kri75}
\begin{eqnarray}
Z_N=\lambda_{1}^{N}+\lambda_{2}^{N}+\lambda_{3}^{N},
\label{1.8}
\end{eqnarray}
which can be evaluated from
\begin{eqnarray}
\lambda_k = \frac{1}{3} \left(u_k g+\frac{a^2-3b}{u_k g}-a \right ).
 \label{1.8a}
\end{eqnarray}
The coefficients $u_k$, $g$, $a$, $b$ and $c$ entering the eigenvalues (\ref{1.8a}) are given by
$\nonumber
\\
u_1=1, u_{2,3}=(-1 \pm i \sqrt{3})/2, \nonumber
\\ g=2^{-1/3}(9ab-2a^3-27c+3\sqrt{3}\sqrt{4b^3-a^2 b^2+4a^3c-18 a b c + 27c^2})^{1/3},
\nonumber
\\
a=-T_{1,1}-T_{0,0}-T_{-1,-1}, \nonumber
\\
b=-T_{1,0}^2-T_{1,-1}^2-T_{-1,0}^2+T_{1,1} T_{0,0} + T_{1,1}
T_{-1,-1}+T_{0,0} T_{-1,-1},\nonumber
\\
c=-2 T_{1,0}T_{1,-1}T_{-1,0}+T_{1,1}T_{-1,0}^2+T_{1,-1}^2T_{0,0}
+T_{1,0}^2T_{-1,-1}-T_{1,1}T_{0,0}T_{-1,-1}.
$

Next, let us denote the largest transfer-matrix eigenvalue $\lambda=
\mbox{max} \{ \lambda_1, \lambda_2,\lambda_3 \}$, because the
contribution of two smaller transfer-matrix eigenvalues to the
partition function may be completely neglected in the thermodynamic
limit $N \rightarrow \infty$
\begin{eqnarray}
Z_N \simeq \lambda^{N}. \label{1.9}
\end{eqnarray}
The free energy per elementary diamond cell can be obtained from the
largest eigenvalue of the transfer matrix (\ref{1.6}) according to
the formula
\begin{eqnarray}
f=- \frac{1}{\beta} \lim_{N \to \infty} \frac{1}{N} \ln Z_N =
-\frac{1}{\beta}\ln\lambda. \label{1.10}
\end{eqnarray}
Knowledge of the free energy allows us to obtain thermodynamic
quantities of the system (such as entropy, magnetization, specific
heat, etc.) in terms of the free energy and/or its derivatives. In
particular, for the entropy $s$ and the heat capacity $c$ per unit
cell one can obtain
\begin{eqnarray}
s=k_B \beta^2 \frac{\partial f}{\partial \beta}, \quad c=-\beta
\frac{\partial s}{\partial \beta}. \label{1.11}
\end{eqnarray}
One may also obtain the single-site magnetization of the Ising
($m_{\mathrm{I}}= \langle \mu_i+\mu_{i+1} \rangle /2$) and the
Heisenberg ($m_{\mathrm{H}}=\langle S_{a,i}^{z}+S_{b,i}^{z}\rangle
/2$) spins, which are given by
\begin{eqnarray}
m_{\mathrm{I}}=-\frac{\partial f}{\partial H_{\mathrm{I}}}, \quad
m_{\mathrm{H}}=-\frac{\partial f}{\partial H_{\mathrm{H}}}.
\label{1.12}
\end{eqnarray}
Finally, the total magnetization per site follows from
\begin{eqnarray}
m=\frac{1}{3}m_\mathrm{I}+\frac{2}{3}m_\mathrm{H}. \label{1.13}
\end{eqnarray}

\section{Results and discussion}
\label{result}

In this section, we will investigate in detail the ground state, magnetization process and specific
heat of the spin-1 Ising-Heisenberg diamond chain with antiferromagnetic coupling constants $J>0$ and $J_1>0$.
Hereafter, we will consider for simplicity the uniform external magnetic field acting
on the Ising and Heisenberg spins $H_{\mathrm{H}}=H_{\mathrm{I}}\equiv H$. Again for simplicity,
the Boltzmann's constant is set to unity $k_B=1$ and a strength of the Ising coupling $J_1$
will be subsequently used for introducing a set of dimensionless parameters
\begin{eqnarray}
\alpha= \frac{J}{J_1}, \quad \gamma= \frac{J_2}{J_1},  \quad
h=\frac{H}{J_1}, \quad t=\frac{T}{J_1},
 \label{2.1}
\end{eqnarray}
which determine a relative strength of the Heisenberg coupling, the second-neighbor interaction between the nodal spins,
the magnetic field and temperature, respectively. The spin-1 Ising-Heisenberg diamond chain without
the second-neighbor interaction between the nodal spins was studied in some detail in our previous work \cite{ans14},
so our particular attention will be primarily devoted to the effect of this interaction term.

Let us at first consider the ground-state phase diagrams of the spin-1 Ising-Heisenberg diamond chain
in zero and non-zero magnetic field. Depending on a relative strength of the second-neighbor interaction
$\gamma$ only two or three different ground states are available at zero magnetic field: the
classical ferrimagnetic (FRI) phase and two quantum antiferromagnetic ones QAF$_1$ and QAF$_2$.
The three aforementioned phases can be characterized by the following eigenvectors,
the ground-state energy and single-site magnetizations:

\begin{itemize}
\item The classical ferrimagnetic phase FRI:
\begin{eqnarray}
|FRI\rangle = \prod^{N}_{i=1}|-1 \rangle_{i} \otimes |1,1 \rangle_{a_i,b_i}, \nonumber\\
E=J+J_2-4J_1-H, m_\mathrm{I}=-1, m_\mathrm{H}=1, m=1/3. \label{2.2a}
\end{eqnarray}

\item The quantum antiferromagnetic phases QAF$_1$ and QAF$_2$:
\begin{eqnarray}
|QAF_1\rangle= \left\{ \begin{array}{l}
\displaystyle \prod^{N}_{i=1}|1 \rangle_{i} \otimes \frac{1}{\sqrt{2}}(|0,-1\rangle-|-1,0\rangle)_{a_i,b_i}, \\
\displaystyle \prod^{N}_{i=1}|-1 \rangle_{i} \otimes
\frac{1}{\sqrt{2}}(|1,0\rangle-|0,1\rangle)_{a_i,b_i},
                        \end{array} \right. \nonumber \\
E=-J \Delta + J_2 -2J_1, m_\mathrm{I}=\pm1, m_\mathrm{H}=\mp0.5,
m=0,
\nonumber \\
|QAF_2\rangle=\prod^{N}_{i=1}|(-1)^i \mathrm{or} (-1)^{i+1}
\rangle_{i} \otimes \frac{\Delta \sqrt{2}}{\sqrt{\delta
(\delta-1)}}(\frac{1-\delta}{2\Delta}|0,0\rangle
\nonumber \\ + |1,-1\rangle+|-1,1\rangle)_{a_i,b_i},\nonumber\\
E=-\frac{J}{2}(1+\delta)-J_2, m_\mathrm{I}=0, m_\mathrm{H}=0, m=0.
\label{2.2b}
\end{eqnarray}
\end{itemize}

\begin{figure}[!h]
\begin{center}
\begin{tabular}{cccc}
{\small (a)}& {\small (b)}\\
\includegraphics[width=6.2cm]{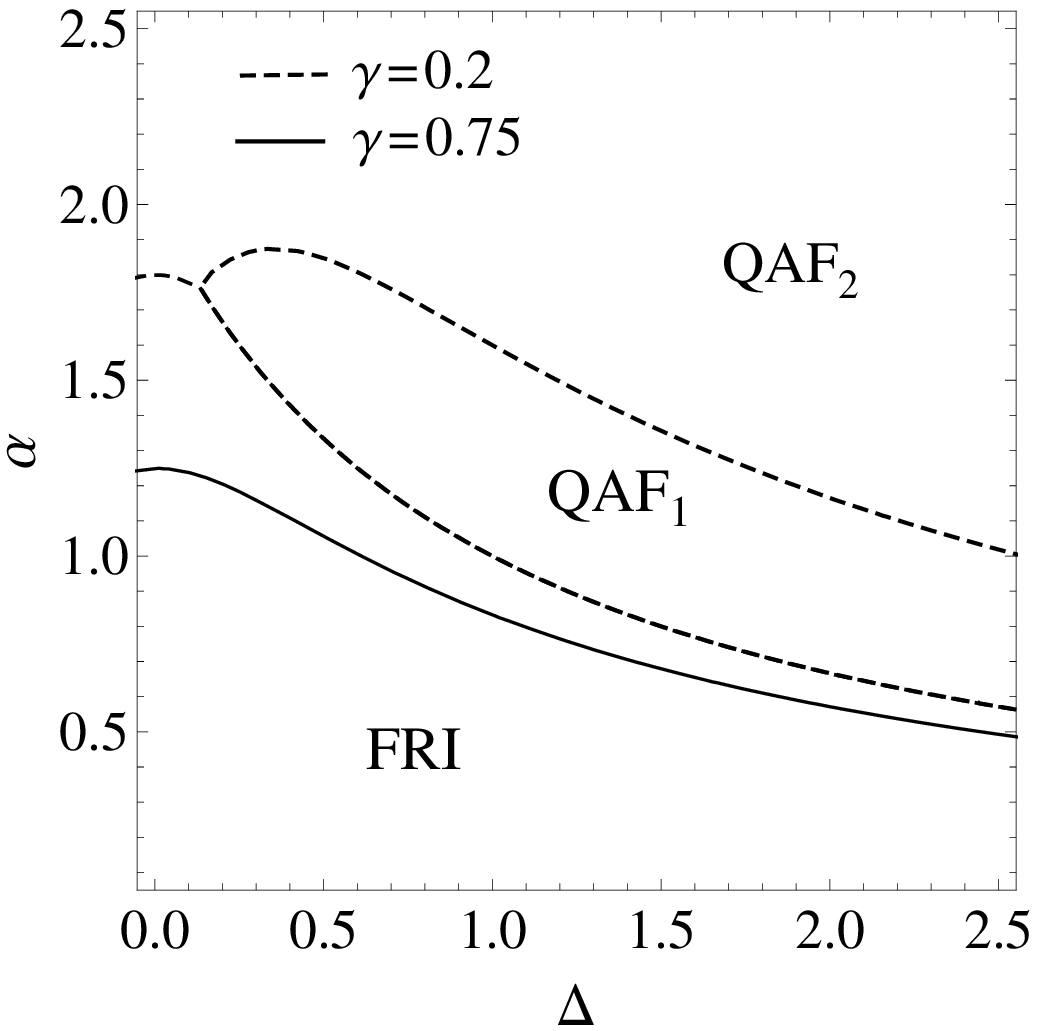} &
\includegraphics[width=6cm]{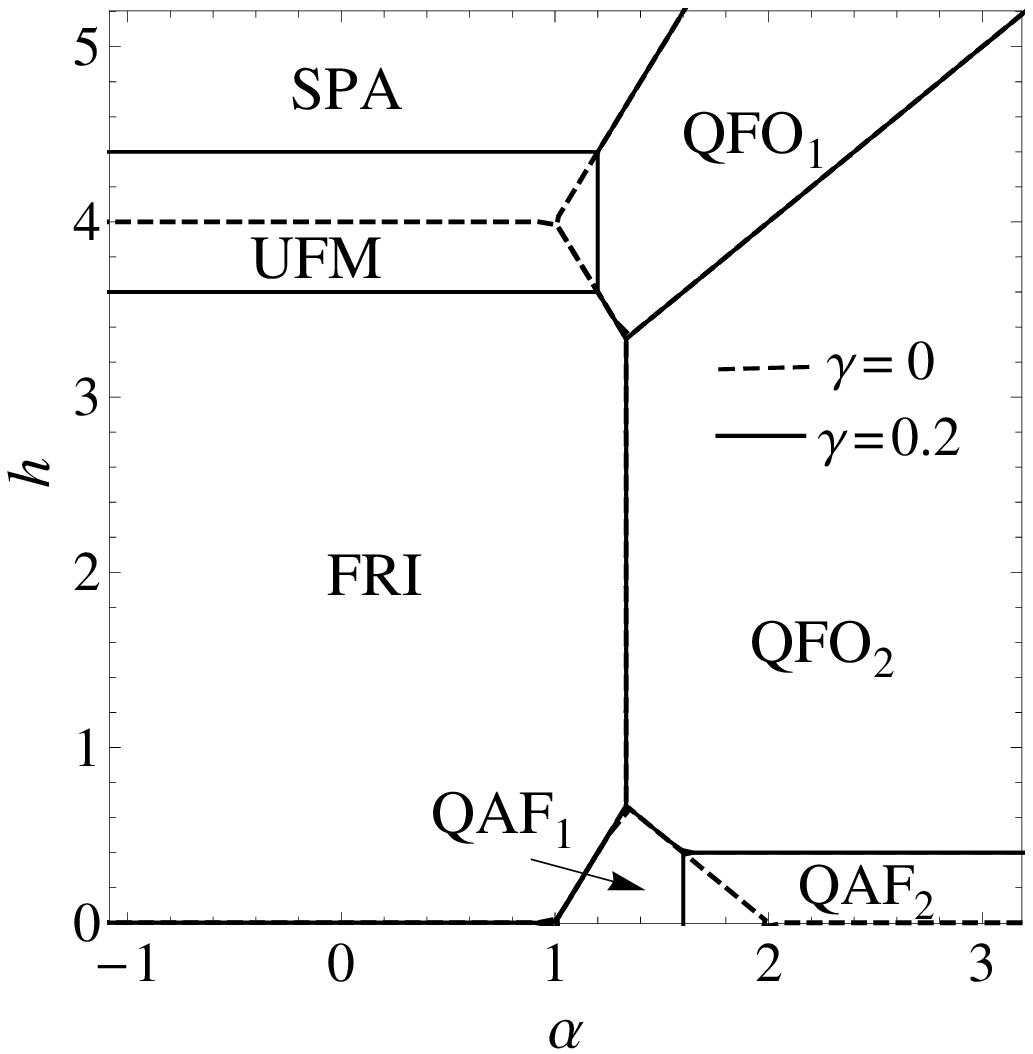}
\nonumber \\
{\small (c)}& {\small (d)}\\
\includegraphics[width=6cm]{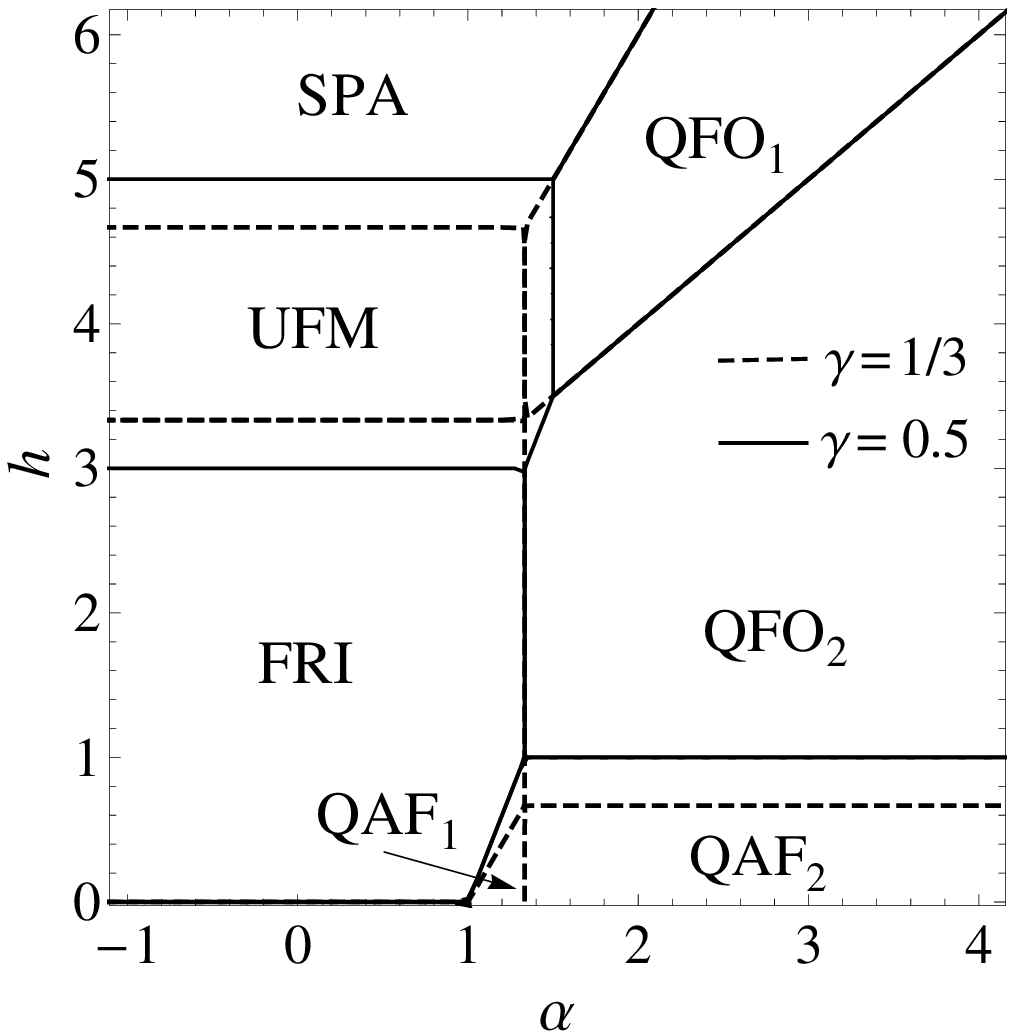} &
\includegraphics[width=6cm]{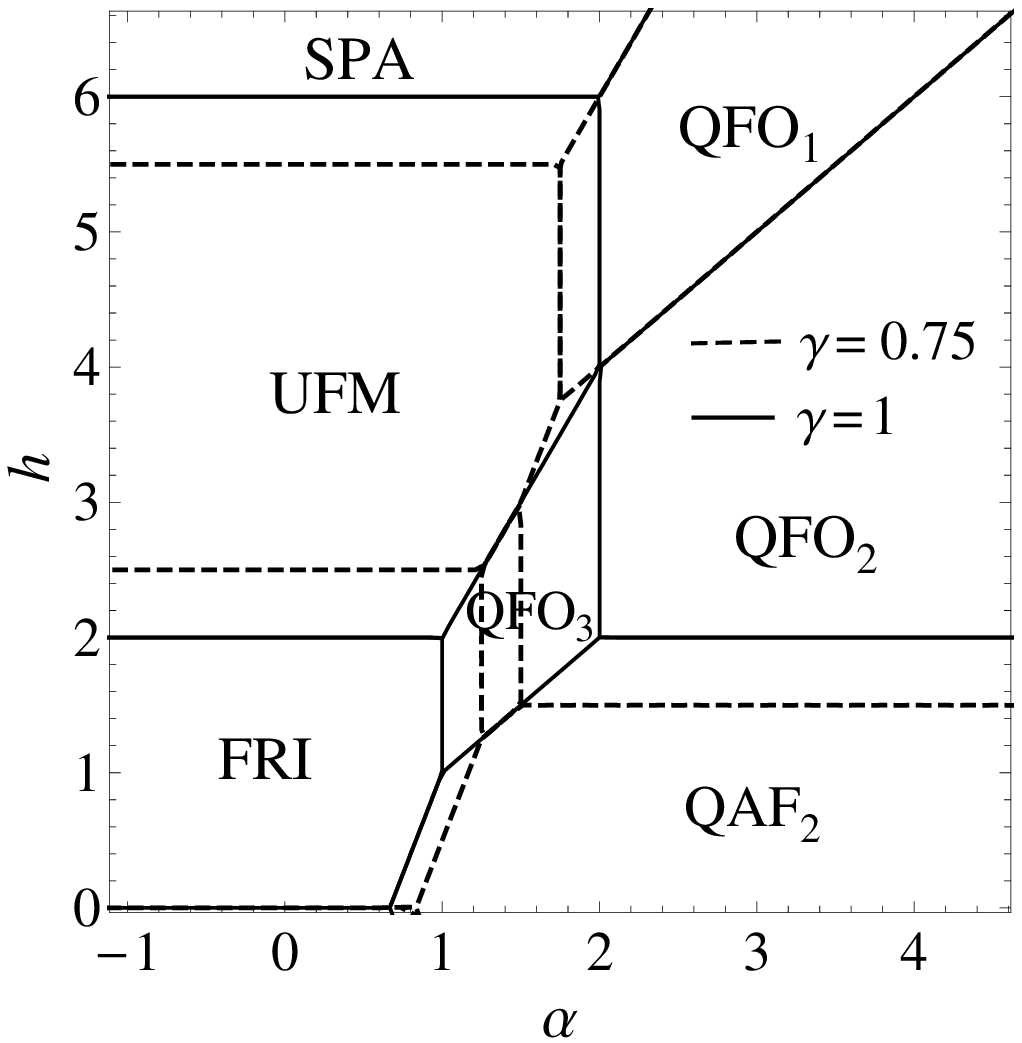}
\end{tabular}
\caption { \small{ (a) Zero-field ground-state phase diagram in the
$\Delta-\alpha$ plane for a few different values of the interaction
parameter $\gamma$; (b)-(d) Ground-state phase diagrams in the
$\alpha-h$ plane for a fixed value of the anisotropy parameter
$\Delta=1$ and a few selected values of the second-neighbor
interaction: (b) $\gamma=0, \gamma=0.2$, (c) $\gamma=1/3,
\gamma=0.5$ and (d) $\gamma=0.75, \gamma=1$.}} \label{Phase}
\end{center}
\end{figure}

The zero-field ground-state phase diagram of the spin-1
Ising-Heisenberg diamond chain is plotted in Fig. \ref{Phase}(a) in
$\Delta-\alpha$ plane for two different values of the
second-neighbor interaction $\gamma$. Note that an increase of the
Heisenberg coupling $\alpha$ strengthens a spin frustration due to a
competition between the antiferromagnetic Heisenberg and Ising
interactions. If the second-neighbor interaction between the nodal
spins is sufficiently small (e.g. $\gamma = 0.2$), then, one passes
from FRI ground state through QAF$_1$ ground state up to QAF$_2$
ground state as the frustration parameter $\alpha$ strengthens. On
the other hand, QAF$_1$ ground state without translationally broken
symmetry is totally absent in the ground-state phase diagram for
strong enough second-neighbor interaction (e.g. $\gamma = 0.75$).

In a presence of the external magnetic field one may additionally
find another five ground states of the spin-1 Ising-Heisenberg
diamond chain: three quantum ferromagnetic (QFO$_1$, QFO$_2$,
QFO$_3$), the unsaturated ferromagnetic (UFM) and the saturated
paramagnetic (SPA), which can be characterized through the following
eigenvectors, the ground-state energy and single-site
magnetizations:

\begin{itemize}
\item The quantum ferromagnetic phases QFO$_1$, QFO$_2$ and QFO$_3$:
\begin{eqnarray}
|QFO_1\rangle=\prod^{N}_{i=1}|1 \rangle_{i} \otimes \frac{1}{\sqrt{2}}(|1,0\rangle-|0,1\rangle)_{a_i,b_i}, \nonumber\\
E=-J \Delta +2J_1 + J_2 -2H, m_\mathrm{I}=1, m_\mathrm{H}=1/2,
m=2/3,
\nonumber \\
|QFO_2\rangle=\prod^{N}_{i=1}|1  \rangle_{i} \otimes \frac{\Delta
\sqrt{2}}{\sqrt{\delta (\delta-1)}}
(\frac{1-\delta}{2\Delta}|0,0\rangle
 + |1,-1\rangle+|-1,1\rangle )_{a_i,b_i},\nonumber\\
E=-\frac{J}{2}(1+\delta)+J_2-H, m_\mathrm{I}=1, m_\mathrm{H}=0,
m=1/3,
\nonumber \\
|QFO_3\rangle=\prod^{N}_{i=1}|(-1)^i \mathrm{or} (-1)^{i+1} \rangle_{i} \otimes \frac{1}{\sqrt{2}}(|1,0\rangle-|0,1\rangle)_{a_i,b_i}, \nonumber\\
E=-J \Delta - J_2 - H, m_\mathrm{I}=0, m_\mathrm{H}=1/2, m=1/3.
 \label{2.3a}
\end{eqnarray}

\item The unsaturated ferromagnetic phase UFM:
\begin{eqnarray}
|UFM\rangle = \prod^{N}_{i=1}|(-1)^i \mathrm{or} (-1)^{i+1} \rangle_{i} \otimes |1,1 \rangle_{a_i,b_i}, \nonumber\\
E=J-J_2-2H, m_\mathrm{I}=0, m_\mathrm{H}=1, m=2/3. \label{2.3b}
\end{eqnarray}

\item The saturated paramagnetic phase SPA:
\begin{eqnarray}
|SPA\rangle=\prod^{N}_{i=1}|1 \rangle_{i} \otimes |1,1 \rangle_{a_i,b_i}, \nonumber\\
E=J+4J_1 + J_2 -3H, \; m_\mathrm{I}=1, m_\mathrm{H}=1, m=1.
\label{2.3c}
\end{eqnarray}
\end{itemize}

The ground-state phase diagram of the spin-1 Ising-Heisenberg
diamond chain in $\alpha-h$ plane is plotted in Fig.
\ref{Phase}(b)-(d) for the isotropic Heisenberg coupling $\Delta=1$
and several values of the second-neighbor interaction $\gamma$. It
can be seen from Fig. \ref{Phase}(b) that the second-neighbor
interaction between the nodal spins gives rise to two new ground
states QAF$_2$ and UFM, which are absent in the model without this
interaction term. In general, the parameter space inherent to the
ground states QAF$_2$ and UFM extends upon rising the
second-neighbor interaction $\gamma$ as evidenced by from Fig.
\ref{Phase}(b) and (c). However, the third novel ground state
QFO$_3$ appears at moderate values of the Heisenberg coupling
$\alpha$ and magnetic field $h$ as far as the second-neighbor
interaction $\gamma$ becomes sufficiently strong (see Fig.
\ref{Phase}(d)).

\begin{figure}[t]
\begin{center}
\begin{tabular}{ccc}
{\small (a)}& {\small (b)}\\
\includegraphics[width=6.5cm]{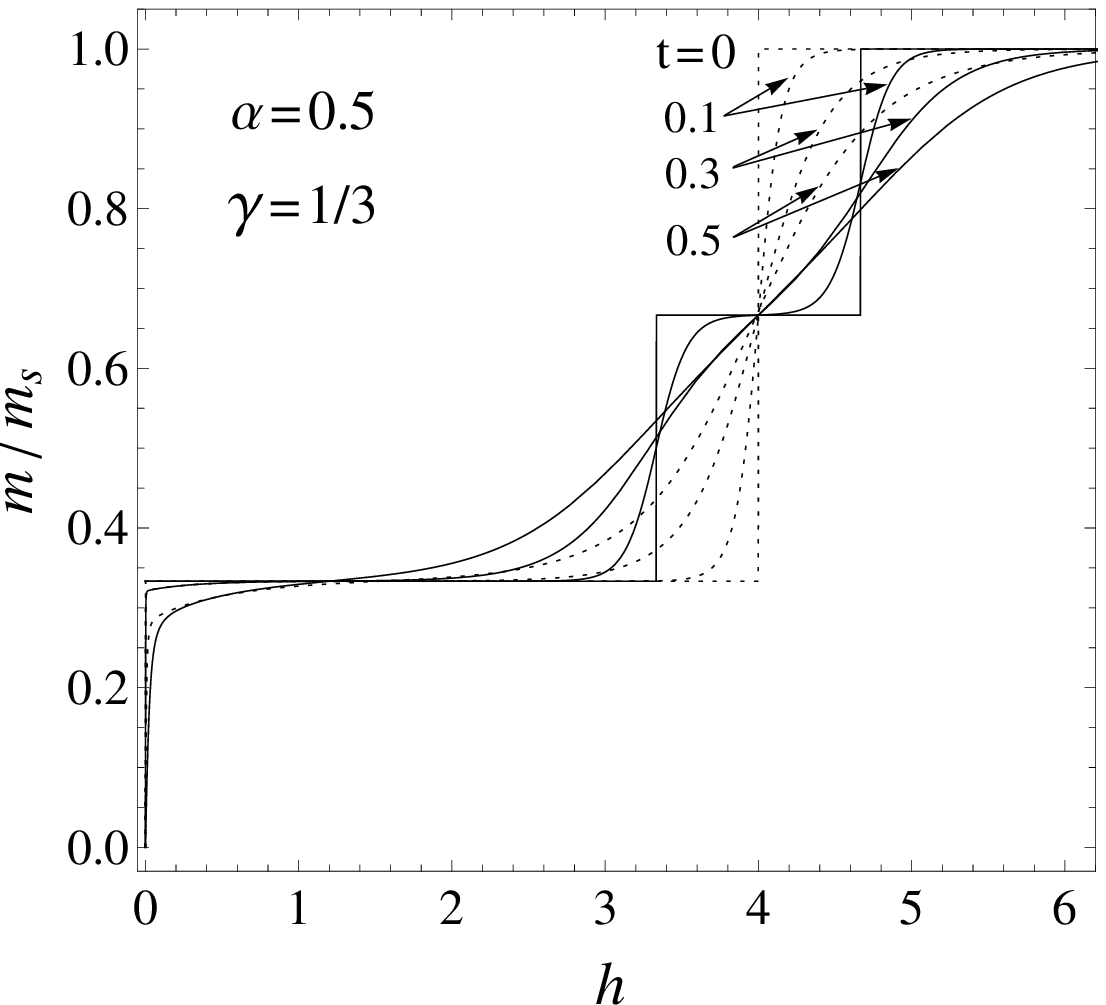}&
\includegraphics[width=6.5cm]{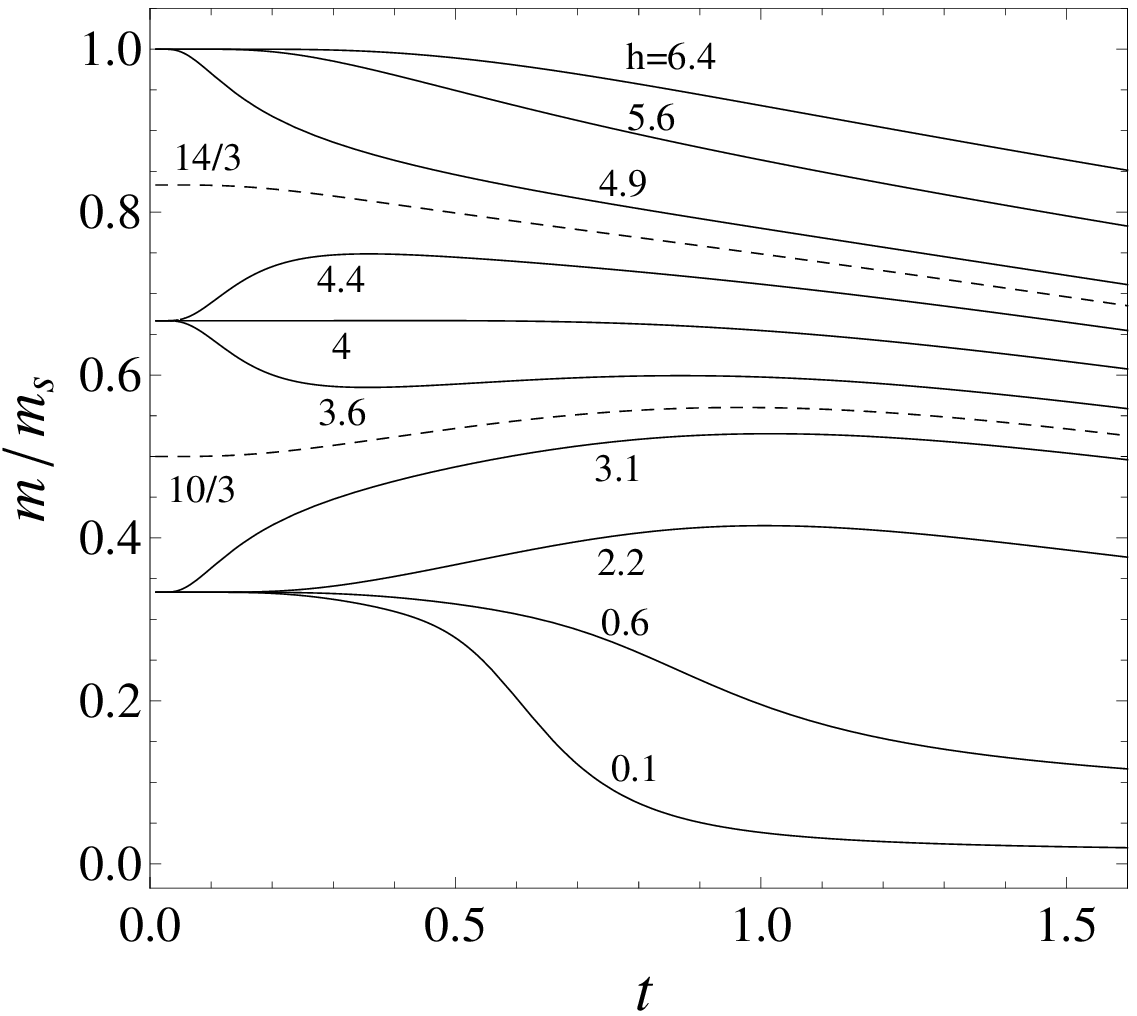}\\
\end{tabular}
\caption {\small{(a) The total magnetization as a function of the
magnetic field at a few different temperatures for the isotropic
Heisenberg interaction ($\Delta=1$) of a relative strength
$\alpha=0.5$. The dotted and solid lines show magnetization curves
for two selected values of the second-neighbor interaction
$\gamma=0$ and $1/3$, respectively. (b) Thermal variations of the
total magnetization for $\Delta=1$, $\alpha=0.5$, $\gamma=1/3$ and
several values of the external magnetic field.}} \label{Mag1}
\end{center}
\end{figure}

\begin{figure}[t]
\begin{center}
\begin{tabular}{ccc}
{\small (a)}& {\small (b)}\\
\includegraphics[width=6.5cm]{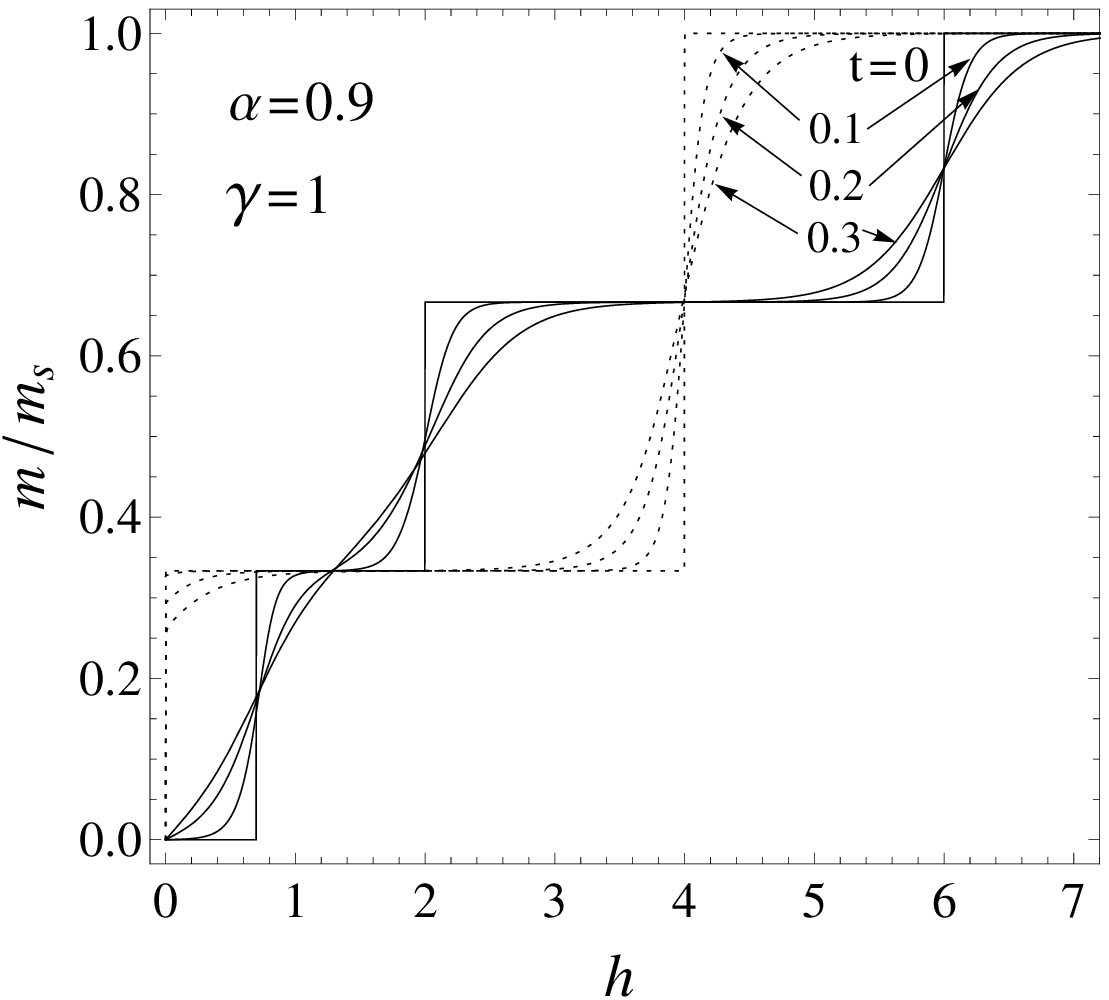}&
\includegraphics[width=6.5cm]{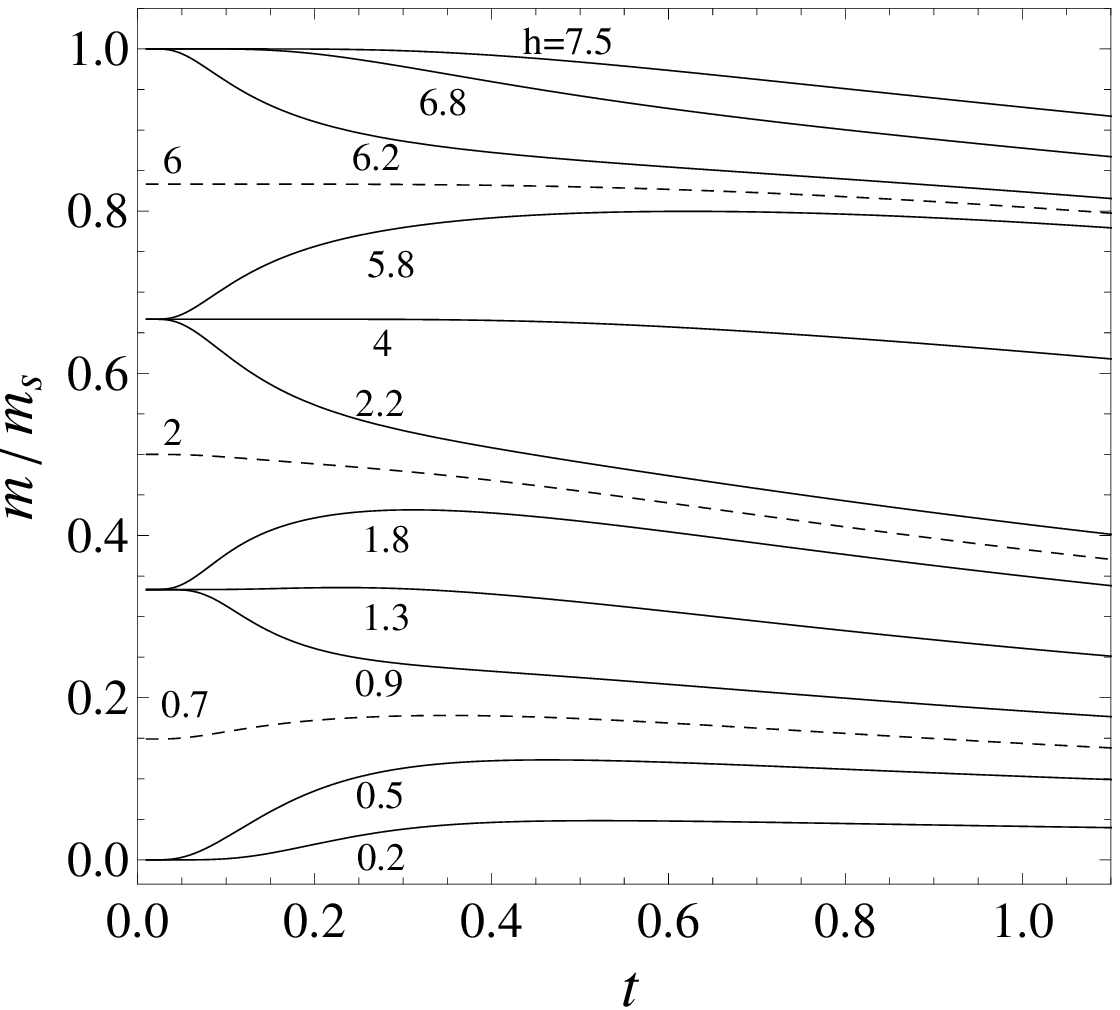}\\
\end{tabular}
\caption {\small{(a) The total magnetization as a function of the
magnetic field at a few different temperatures for the isotropic
Heisenberg interaction ($\Delta=1$) of a relative strength
$\alpha=0.9$. The dotted and solid lines show magnetization curves
for two selected values of the second-neighbor interaction
$\gamma=0$ and $1$, respectively. (b) Thermal variations of the
total magnetization for $\Delta=1$, $\alpha=0.9$, $\gamma=1$ and
several values of the external magnetic field.}} \label{Mag2}
\end{center}
\end{figure}

Now, let us proceed to a comprehensive analysis of the magnetization
process at zero as well as non-zero temperatures. To illustrate an
influence of the second-neighbor coupling on a magnetization
process, Figs. \ref{Mag1}(a) and \ref{Mag2}(a) compare two different
magnetization scenarios of the spin-1 Ising-Heisenberg diamond chain
with and without this interaction term. Fig. \ref{Mag1}(a) displays
the magnetization curve with a single one-third plateau due to a
field-induced transition FRI-SPA for $\gamma=0$ along with the
magnetization curve with two successive one-third and two-thirds
plateaus, which emerge due to two subsequent field-induced
transitions FRI-UFM-SPA for $\gamma=1/3$. Interestingly, the simple
magnetization curve with a single one-third plateau due to a
field-induced transition FRI-SPA for $\gamma=0$ may also change to a
more complex magnetization curve with three successive plateaus at
zero, one-third and two-thirds of the saturation magnetization,
which emerge due to three subsequent field-induced transitions
QAF$_2$-FRI-UFM-SPA for $\gamma=1$. It is worthwhile to recall that
actual magnetization plateaus and magnetization jumps only appear at
zero temperature, while increasing temperature gradually smoothens
the magnetization curves. Typical thermal variations of the total
magnetization are plotted Figs. \ref{Mag1}(b) and \ref{Mag2}(b),
which evidence pronounced low-temperature variations of the total
magnetization when the magnetic field is fixed slightly below or
above the relevant critical field.

\begin{figure}[t]
\begin{center}
\begin{tabular}{ccc}
{\small (a)}&{\small (b)}\\
\includegraphics[width=6cm]{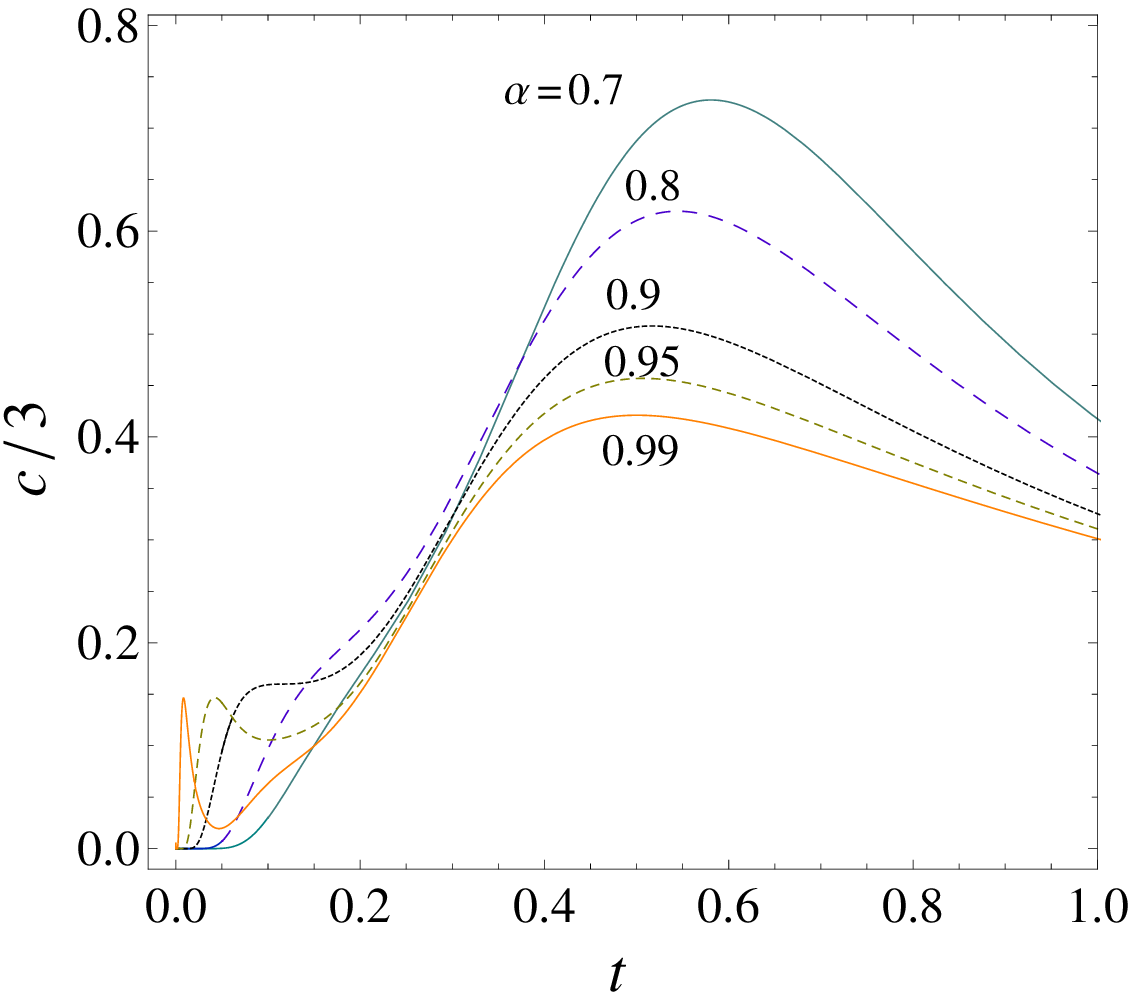}&
\includegraphics[width=6cm]{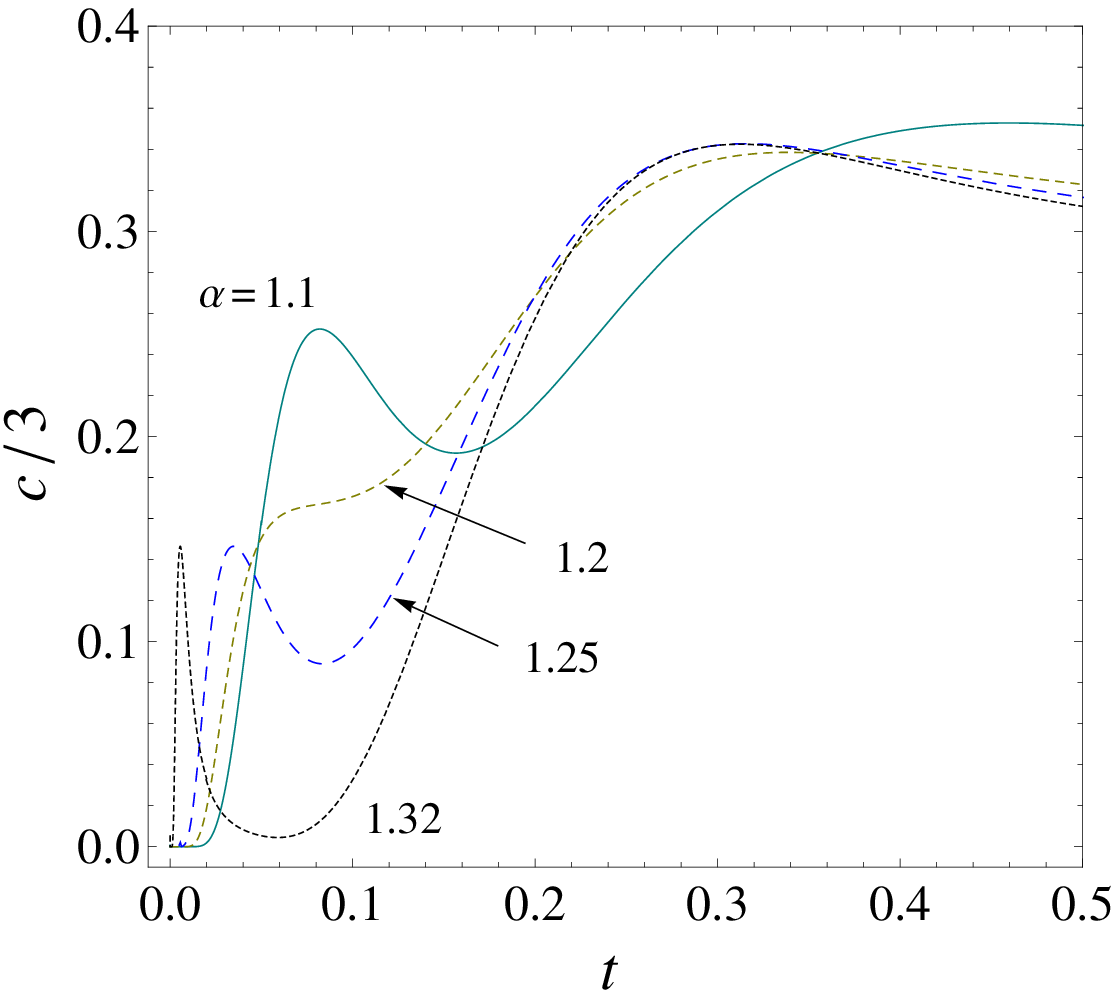}
\nonumber \\
{\small (c)}\\
\includegraphics[width=6cm]{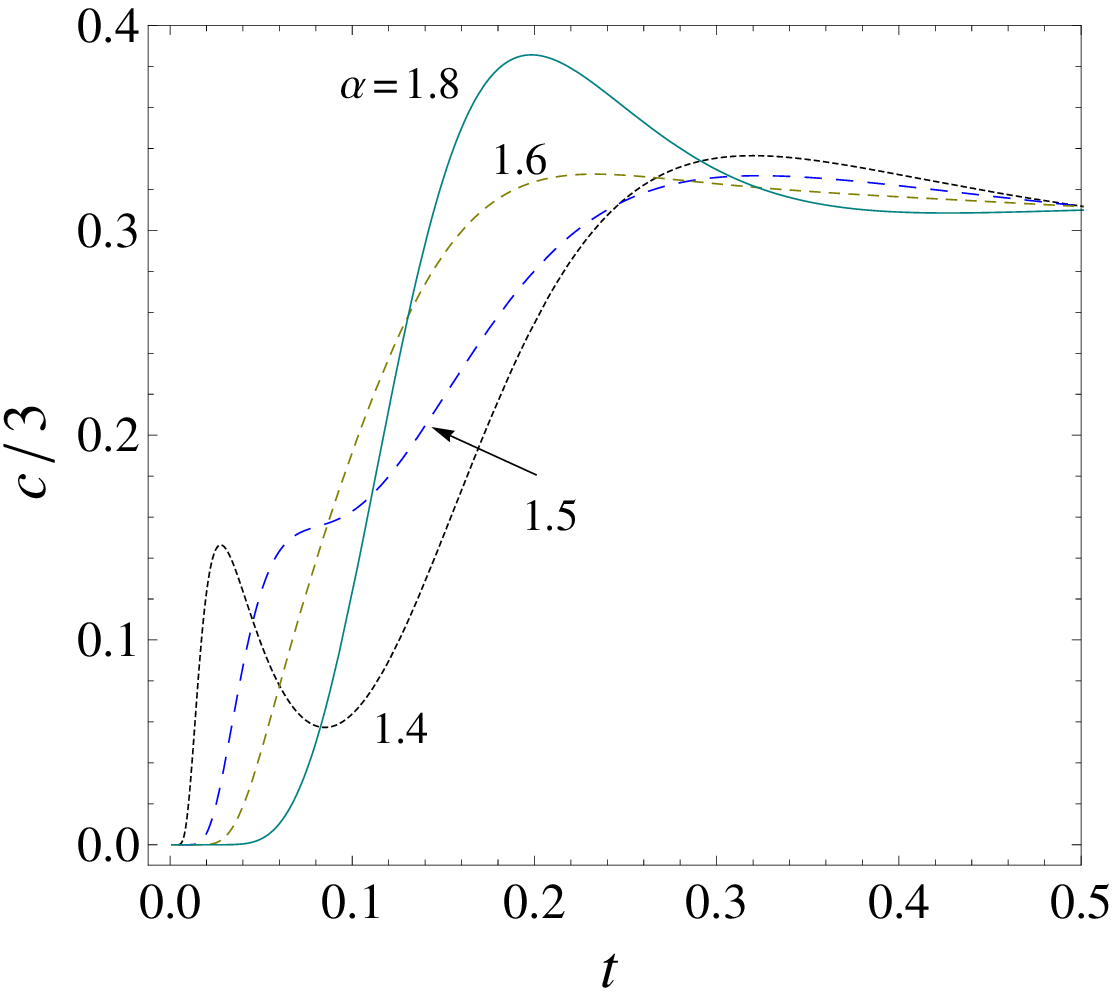}\\
\end{tabular}
\caption {\small{Typical temperature dependences of the zero-field specific heat
for the fixed value of the second-neighbor interaction $\gamma=1/3$ and a few
different values of the isotropic Heisenberg coupling with $\Delta=1$.
Three panels correspond to three available zero-field ground states:
(a) FRI, (b) QAF$_1$ and (c) QAF$_2$.}} \label{SHeat}
\end{center}
\end{figure}

Last but not least, let us briefly comment on thermal variations of
the zero-field specific heat, which are quite typical for three
available zero-field ground states FRI,  QAF$_1$ and  QAF$_2$.
Temperature dependencies of the zero-field specific heat pertinent
to the classical FRI ground state are depicted in Fig.
\ref{SHeat}(a). It can be seen from this figure that the specific
heat may display a more peculiar double-peak temperature dependence
in addition to a standard temperature dependence with a single round
maximum. The double-peak temperature dependencies of the specific
heat are also quite typical for another ground state QAF$_1$, which
appears in a rather restricted region of parameter space (see Fig.
\ref{SHeat}(b)). On assumption that the QAF$_2$ phase constitutes
the ground state thermal variations of the zero-field specific heat
with a single- or double-peak structure emerge for greater or
smaller values of the Heisenberg interaction $\alpha$ as depicted in
Fig. \ref{SHeat}(c). It could be concluded that the double-peak
temperature dependencies of the specific heat appear in all three
aforementioned cases owing to low-lying thermal excitations and
consequently, the low-temperature peak can be always identified as
the Schottky-type maximum.

\section{Conclusion}
\label{conclusion}

In the present work, we have examined the ground state,
magnetization process and specific heat of the exactly solved
spin-$1$ Ising-Heisenberg diamond chain with the second-neighbor
interaction between the nodal spins. It has been demonstrated that
the considered further-neighbor interaction gives rise to three
novel ground states, which cannot be in principle detected in the
simplified version of the spin-$1$ Ising-Heisenberg diamond chain
without this interaction term \cite{ans14}. It should be pointed
out, moreover, that the spin-$1$ Ising-Heisenberg diamond chain
supplemented with the second-neighbor interaction between the nodal
spins does not exhibit more intermediate magnetization plateaus than
its simplified version without this interaction term even though all
three novel ground states have translationally broken symmetry. This
finding seems to be quite surprising, because one could generally
expect according to Oshikawa-Yamanaka-Affleck rule
\cite{osh97,aff98} appearance of new intermediate plateaus at
one-sixth and five-sixths of the saturation magnetization in
addition to the observed intermediate plateaus at zero, one-third
and two-thirds of the saturation magnetization. Of course, one
cannot definitely rule out that the one-sixth and/or five-sixths
plateaus indeed occur in a zero-temperature magnetization curve of a
more general spin-$1$ Ising-Heisenberg diamond chain, which could
for instance take into account asymmetric interactions, single-ion
anisotropy, four-spin and/or biquadratic interactions besides the
second-neighbor interaction between the nodal spins. This conjecture
might represent challenging task for future investigations.

\section*{Acknowledgments}

J S acknowledges financial support provided by The Ministry of
Education, Science, Research, and Sport of the Slovak Republic under
Contract Nos. VEGA 1/0234/12 and VEGA 1/0331/15 and by grants from
the Slovak Research and Development Agency under Contract Nos.
APVV-0097-12 and APVV-14-0073. N A acknowledges financial support by
the MC-IRSES no. 612707 (DIONICOS) under FP7-PEOPLE-2013 and
research project no. SCS 15T-1C114 grants.

\end{document}